\documentclass[namedreferences]{solarphysics}
\usepackage[hyperref,optionalrh,showbiblabels]{spr-sola-addons} 
\usepackage{graphicx}                    
\usepackage{amssymb}                     
\usepackage{color}                       
\usepackage{url}                         
\usepackage{breakurl}                    
\usepackage[utf8]{inputenc}
\usepackage[T1]{fontenc}
\usepackage{textcomp}
\usepackage{siunitx}
\usepackage{tabularx,booktabs}
\usepackage{adjustbox}
\usepackage{lipsum}
\usepackage{float}
\usepackage{enumerate}
\usepackage{multirow}
\usepackage{caption}
\usepackage{threeparttablex}
\usepackage{ulem}
\usepackage{xfrac}

\begin{document}

\begin{article}

\begin{opening}

\title{Flare Productivity of Major Flaring Solar Active Regions: A Time-series Study of Photospheric Magnetic Properties}


\author[addressref=aff1,email={eo-jin.lee@khu.ac.kr}]{\inits{E.-J.}\fnm{Eo-Jin}~\lnm{Lee}}
\author[addressref=aff2,corref,email={shpark@isee.nagoya-u.ac.jp}]{\inits{S.-H.}\fnm{Sung-Hong}~\lnm{Park}\orcid{0000-0001-9149-6547}}
\author[addressref={aff1,aff3},email={moonyj@khu.ac.kr}]{\inits{Y.-J.}\fnm{Yong-Jae}~\lnm{Moon}}


\address[id=aff1]{School of Space Research, Kyung Hee University, Yongin, Republic of Korea}
\address[id=aff2]{Institute for Space-Earth Environmental Research, Nagoya University, Nagoya, Japan}
\address[id=aff3]{Department of Astronomy and Space Science, Kyung Hee University, Yongin, Republic of Korea}


\runningauthor{Lee et al.}
\runningtitle{Flare Productivity and Magnetic Properties of Flaring Active Regions}


\begin{abstract}
A solar active region (AR) that produces at least one M- or X-class major flare tends to produce multiple flares during its passage across the solar disk. It will be interesting if we can estimate how flare-productive a given major flaring AR is for a time interval of several days, by investigating time series of its photospheric magnetic field properties. For this, we studied 93 major flaring ARs observed from 2010 to 2016 by the Helioseismic and Magnetic Imager (HMI) on board the Solar Dynamics Observatory (SDO). More specifically, for each AR under study, the mean and fluctuation were calculated from an 8-day time series of each of 18 photospheric magnetic parameters extracted from the Space-weather HMI Active Region Patch (SHARP) vector magnetogram products at 12-min cadence. We then compared these with the AR's 8-day flare index, which is defined as the sum of soft X-ray peak fluxes of flares produced in the AR during the same interval of the 8-day SHARP parameter time series. As a result, it is found that the 8-day flare index is well correlated with the mean and/or fluctuation values of some magnetic parameters (with correlation coefficients of 0.6--0.7 in log-log space). Interestingly, the 8-day flare index shows a slightly better correlation with the fluctuation than the mean for the SHARP parameters associated with the surface integral of photospheric magnetic free energy density. We also discuss how the correlation varies if the 8-day flare index is compared with the mean or fluctuation calculated from an initial portion of the SHARP parameter time series.
\end{abstract}



\end{opening}

\section{Introduction}
\label{s:int}
Solar flares suddenly release a huge amount of energy mainly in the form of electromagnetic radiation and high-energy particles. Consequently, flares occasionally cause rapid, significant variations in the near-Earth space environment called space weather \citep[e.g.,][]{2014SpWea..12..205H,2017JGRA..122.9841H}. There have been several reports that (severe) space weather events related to major flares affect space assets and also harm human health, such as malfunction of satellites, radiation exposure of astronauts/aircrews/passengers, interruption of radio communication, and so on \citep[e.g.,][]{Baker2005,2014SpWea..12..487S,Lugaz2015,2015SPIE.9680E..5GP}. Various efforts are therefore being made to establish a reliable system for operational flare prediction to effectively prevent various types of damage from powerful flare events. \citep[e.g., refer to][and several flare forecasting methods therein]{2016ApJ...829...89B}

In order to understand energy build-up and triggering processes of flares in solar active regions (ARs), it is very important to investigate three-dimensional AR magnetic field structures and their evolution. However, direct measurement of the coronal magnetic field has been scarcely carried out, so that photospheric magnetic field data have been mainly used to study various AR magnetic field properties in relation to flares \citep[e.g., refer to][and references therein]{2003ApJ...595.1296L}. For example, \citet{1987SoPh..113..347M} studied the shear angle (i.e., the angular deviation of the observed transverse magnetic field from the potential transverse field) along the magnetic polarity inversion lines (PILs) in three different ARs, and they found that the three large flares under investigation occurred in extended PIL regions with large magnetic shear. Measuring the total unsigned magnetic flux around strong-gradient PILs of ARs (called log \textit{R} or R\_VALUE), \citet{2007ApJ...655L.117S} found that the larger R\_VALUE an AR has, the higher chance it produces a flare within the next 24 hours following the R\_VALUE measurement. \citet{2007ApJ...656.1173L} carried out some statistical tests based on linear discriminant analysis with numerous magnetic parameters derived from $\sim$1200 photospheric vector magnetograms of 496 different ARs. They found that the best-performing discriminant functions, resulting from combining three or more photospheric magnetic parameters, make a slight improvement to distinguish between flaring and flare-quiet ARs. Recently, several machine-learning algorithms, such as support vector machine and multilayer perception, have been applied to various AR photospheric magnetic parameters in order to improve the performance of flare prediction \citep[e.g.,][]{2013SoPh..283..157A,2015ApJ...798..135B,2017ApJ...843..104L,2017ApJ...849..148S}.

An AR producing at least one M- or X-class major flare, called a major flaring AR, in general consists of several, large sunspots with strong magnetic fields, and it often has a complicated magnetic field structure which may evolve dynamically in the form of rotation, shear motion or flux emergence/cancellation \citep[e.g.,][]{2001ApJ...548L..99Z,2008ApJ...686.1397P,2009SoPh..258..203M,2014SoPh..289.3351T,2014ApJ...787...88Z,2016ApJ...828...62J}. Major flaring ARs tend to produce multiple flares of different sizes, typically accompanied with high-speed and geoeffective CMEs, during its solar disk passage \citep[e.g.,][]{2005JGRA..11012S05Y,2009ApJ...703..757L}. There have been few studies to understand how long-term (i.e., several days) flare productivity of major flaring ARs is related to properties and evolution of their photospheric magnetic field. \citet{2010ApJ...720..717A} examined power spectra calculated from line-of-sight magnetograms of 217 flaring ARs that produced at least one flare of any GOES class. They found that a steeper magnetic power spectrum is shown in ARs with higher flare productivity during their solar disk passage. Note that the power spectrum of each AR is calculated from the AR's magnetogram observed at a single point in time of its passage near the solar disk center. \citet{2009ApJ...705..821W} determined several time-averaged magnetic and flow field parameters calculated from $\sim$3--5-day time series of line-of-sight magnetograms of 46 ARs. Comparing the parameters to flare productivity of the ARs during the same time interval of the investigated magnetogram time series, they found that in general the larger the value of some parameters an AR has, the higher flare-productive it is. Note that among the 46 ARs, only 11 of them are major flaring ones.

In the present study, we investigate flare productivity of major flaring ARs in relation to their photospheric magnetic field properties with the following important differences compared to previous studies. (1) We investigate a large number of major flaring ARs, i.e., 93 different NOAA-numbered ARs with at least one M- or X-class flare, observed between August 2010 and February 2016. (2) We examine 8-day time series of various parameters derived from AR photospheric vector magnetograms at 12-min cadence. (3) The time series data is parameterized with the mean and fluctuation. (4) We study how the mean or fluctuation is correlated with flare productivity during the same interval (i.e., 8 days) of the time series. (5) We discuss how the correlation varies if we use an initial portion of the time series to calculate the mean and fluctuation, but the same 8-day flare productivity.

\section{Data and Analysis}
\label{s:data}
The Helioseismic and Magnetic Imager \citep[HMI;][]{2012SoPh..275..207S} on board the Solar Dynamics Observatory \citep[SDO;][]{2012SoPh..275....3P} provides full-disk, photospheric vector magnetograms with a spatial resolution of 0.5 arcsec per pixel and temporal resolution of 12 minutes. In this study, we use 18 magnetic parameters stored as keywords in ``hmi.sharp\_cea\_720s series'' of Space-weather HMI Active Region Patch \citep[SHARP;][]{2014SoPh..289.3549B} data products. The parameters are derived for each automatically-identified HMI AR patch (called HARP) using its corresponding photospheric vector magnetogram which is deprojected to the heliographic coordinates with a Lambert (cylindrical equal area; CEA) projection method. The SHARP parameters generally characterize the identified AR's strong-magnetic-field area, unsigned magnetic flux in the entire region or around PILs, magnetic field inclination/gradient/shear/twist, current density, current helicity, and magnetic free (excess) energy \citep{2003ApJ...595.1296L,2007ApJ...655L.117S,2014SoPh..289.3549B,2015ApJ...798..135B}. There have been attempts to predict flares and coronal mass ejections using the SHARP parameters \citep{2015ApJ...798..135B, 2016ApJ...821..127B}. Refer to Table~\ref{table_1} for the details of the 18 SHARP parameters.

The noise level in the HMI observables shows large-scale spatial variations over the entire solar disk due irregular characteristics of the HMI instrument \citep{2016SoPh..291.1887C}. \citet{2014SoPh..289.3483H} reported that the number of high-confidence pixels, during the disk passage of HARP 2920, decreases significantly as the HARP moves away from a region, hereafter $\Theta_{60}$, within $\sim$60$^\circ$ from the central meridian. We therefore use the SHARP parameters here, only if they are calculated from an HARP of which center position is within $\Theta_{60}$. For our data set, we aimed to include a large number of major flaring ARs in order to study the relationship between their long-term flare productivity and various photospheric magnetic field properties (i.e., the SHARP parameters). Over the period of HMI observations from August 2010 to February 2016, we searched ARs that: (1) consistently appeared with a well-developed structure within $\Theta_{60}$ so that their associated SHARP data are available, and (2) produced at least one M- or X-class flare during their passage within $\Theta_{60}$. As a result, we found a total of 93 different NOAA-numbered, major flaring ARs. 

For each of the 93 major flaring ARs under study, we analyze a set of time series of the 18 SHARP parameters at 12-min cadence over $\sim$8 days of its passage within $\Theta_{60}$. For all the examined ARs, a time series $\textbf{X}$ of a given, single SHARP parameter consists of a total of 960 data points at 12-min cadence in the selected 8-day interval: i.e., $\textbf{X}=\left[\mathrm{X}_{1},\mathrm{X}_{2},\mathrm{X}_{3},\ldots, \mathrm{X}_{960}\right]$. If there is however a missing value in the 8-day time series, then it is filled with a Not-a-Number (NaN) representing an undefined or unrepresentable value in numeric calculations such as the mean and standard deviation. For the time series $\textbf{X}$, we first calculate the mean as
\begin{equation}
\mathrm{\bar{X}} = \frac{1}{N} \sum_{i=1}^{960} \mathrm{X}_{i},
\end{equation}
where the summation is done excluding NaNs if they exist in $\textbf{X}$ and $N$ is the number of all data points except NaNs. There are three SHARP parameters which are signed quantities: i.e., the mean vertical current density MEANJZD, the mean characteristic twist parameter MEANALP, and the mean vertical current helicity MEANJZH. For these signed parameters, note that we use the absolute value $\left|\mathrm{\bar{X}}\right|$ of the mean of their $\textbf{X}$ for the comparison with 8-day flare productivity of the ARs under study during their disk passage within $\Theta_{60}$. Note that the SHARP parameters are calculated from only pixels that are (1) within the smooth bounding curve (i.e., {\small BITMAP}\,$\geq$\,33) of the HARP rectangular bounding box and (2) above the high-confidence disambiguation threshold (i.e., {\small CONF\_DISAMBIG}=90). The number of the pixels contributing to the SHARP parameter calculation varies with location on the disk and velocity of SDO relative to the Sun \citep{2014SoPh..289.3483H} so that systematic errors, including the center-to-limb variation and the 12-hour periodicity, are embedded in $\textbf{X}$. It is expected that some systematic errors may be reduced by averaging the parameter values in $\textbf{X}$. 

Detrended fluctuation analysis (DFA) is a technique to investigate long-range dependence and statistical self-similarity in both stationary and non-stationary time series \citep[e.g.,][]{1994PhRvE..49.1685P, 1995Chaos...5...82P, 2001PhRvE..64a1114H,2005MAP....88..119K}. It can be also applied to estimate the characteristic size of fluctuation from the cumulative and detrended time series of measurements, such as $\textbf{X}$, consisting of a time-varying signal with random noise and/or a systematic error. This detrended fluctuation primarily captures apparently random variations in $\textbf{X}$ by diminishing variations which are relatively more persistent and/or more systematic. Such random variations in $\textbf{X}$ might be important for self-organized-criticality (SOC) models of flares, in which the corona reaches a critical state and is driven to flare by random motions of photospheric footpoints of coronal loops \citep[e.g.,][]{1991ApJ...380L..89L,2001SoPh..203..321C}. The detrended fluctuation of $\textbf{X}$ in this study is determined as follows. First, the cumulative sum, $\textbf{Y}=\left[\mathrm{Y}_{1},\mathrm{Y}_{2},\mathrm{Y}_{3},\ldots, \mathrm{Y}_{960}\right]$, is calculated from a sequence of $\textbf{X}-\mathrm{\bar{X}}$ as
\begin{equation}
\mathrm{Y}_{k} = \sum_{i=1}^k \left( \mathrm{X}_{i} - \mathrm{\bar{X}} \right).
\end{equation}
Next, the time series $\textbf{Y}$ is divided into non-overlapping segments of equal length $n$, and for each segment a local least squares straight-line fit $\mathrm{\tilde{Y}}(n)$ (i.e., a local linear trend) is calculated. The detrended fluctuation $\mathrm{\tilde{F}}(n)$ is then defined as the root-mean-square deviation of $\textbf{Y}$ with respect to the local trend $\mathrm{\tilde{Y}}(n)$, i.e.,
\begin{equation}
\mathrm{\tilde{F}}(n) = \sqrt{\frac{1}{N} \sum_{i=1}^{960} \left[ \mathrm{Y}_{i} - \mathrm{\tilde{Y}}_{i}(n) \right]^2}.
\end{equation}
$\mathrm{\tilde{F}}(n)$ is one of quantitative measures that can be used to estimate the overall degree of variation in time series data $\textbf{X}$, including short-term and long-term fluctuations as well as linear and non-linear trends but eliminating the local linear trend $\mathrm{\tilde{Y}}(n)$ in the cumulative sum $\textbf{Y}$. We calculated $\mathrm{\tilde{F}}(n)$ with a set of different values of $n$, i.e., $n$=[4, 6, 8, 10, 15, 20, 30, 40, 48, 60, 80, 120], and found that the correlation of $\mathrm{\tilde{F}}(n)$ with the 8-day flare productivity is only weakly sensitive to $n$. In this study, $\mathrm{\tilde{Y}}$ is calculated with $n$=120 (i.e., 24-hr, non-overlapping intervals) in order to reduce any contamination in $\mathrm{\tilde{F}}(n)$ due to the SDO's orbital signal. Hereafter, by fluctuation $\mathrm{\tilde{F}}$, we refer to the detrended fluctuation using $n$=120, i.e., $\mathrm{\tilde{F}}(120)$.

\begin{figure}
\centering
\includegraphics[width=\textwidth]{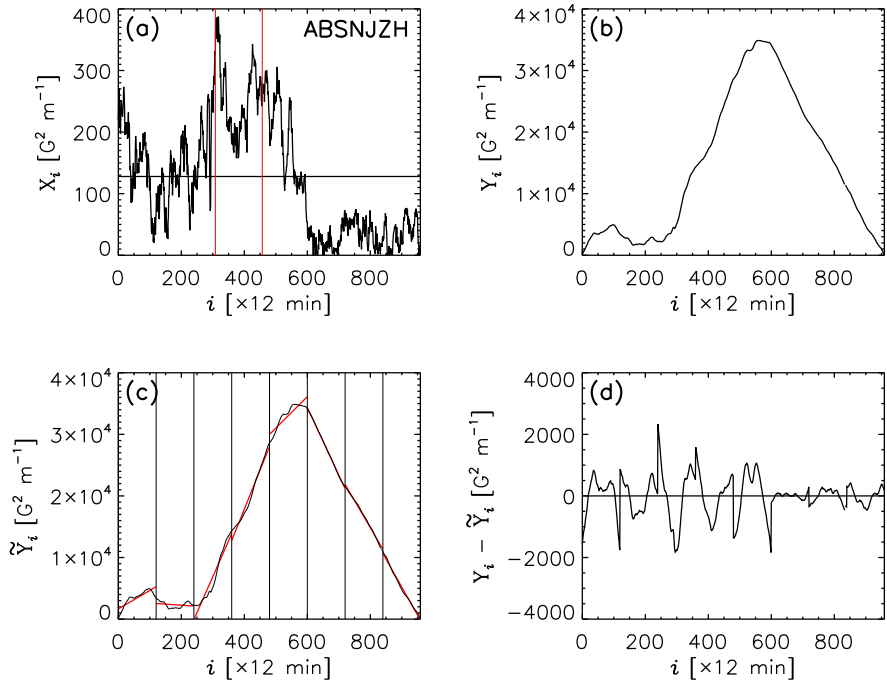}
\caption{An example of calculating the mean $\mathrm{\bar{X}}$ and the detrended fluctuation $\mathrm{\tilde{F}}$. \textit{Panels} show (a) the 8-day time series $\textbf{X}$ (black line) of one of the SHARP parameters under study, i.e., the absolute value of the net vertical current helicity ABSNJZH in HARP 5011 and its $\mathrm{\bar{X}}$ (horizontal line), (b) the cumulative sum $\textbf{Y}$ of $\textbf{X}$, (c) the local least squares straight-line fit $\mathrm{\tilde{Y}}$ for each segment of equal length $n$=120 (red line) on $\textbf{Y}$ (black line), and (d) the deviation from the local trend, i.e., $\textbf{Y}-\mathrm{\tilde{Y}}$. $\mathrm{\tilde{F}}$ is defined as the root-mean-square of $\textbf{Y}-\mathrm{\tilde{Y}}$. The start times of all major flares (M1.0 above), produced in the HARP during the interval of $\textbf{X}$, are marked in the panel \textit{a} with vertical red lines. Note that there are two missing data points in $\textbf{X}$ which are filled with Not-a-Number (NaN) values.}
\label{f1}
\end{figure}

Figure~\ref{f1}{a} presents an example of the time series $\textbf{X}$ of the absolute value of the net vertical current helicity ABSNJZH for HARP 5011, from 31 December 2014 to 8 January 2015. The 12-hour periodicity appears in $\textbf{X}$, which is due to the spacecraft orbital velocity relative to the Sun. The mean value $\mathrm{\bar{X}}$ is marked with the horizontal red line in Fig.~\ref{f1}{a}. The cumulative sum $\textbf{Y}$ is shown in Fig.~\ref{f1}{b}, and the local least squares best-fit line $\mathrm{\tilde{Y}}$ (red) for each segment of $n$=120 is overplotted with $\textbf{Y}$ (black) in Fig.~\ref{f1}{c}. The deviation from the local trend, i.e., $\mathrm{Y}_{i} - \mathrm{\tilde{Y}}_{i}$, is plotted in Fig.~\ref{f1}{d}. We can see the 12-hour periodicity is somewhat reduced in the time profile of the deviation. 

To quantify the 8-day flare productivity of each AR under our study, we define 8-day flare index (hereafter shortly indicated as $I_\mathrm{8d}$) as the sum of GOES soft X-ray peak fluxes of flares produced in the AR during the same interval of the AR's 8-day SHARP parameter time series data, i.e.,
\begin{equation}
I_\mathrm{8d} =  100 \times S^{\left( X \right)} + 10 \times S^{\left( M \right)} + 1 \times S^{\left( C \right)},
\end{equation}
where $S^{\left( j \right)} = \sum_{i=1}^{N_j} \xi_{i}^{\left( j \right)}$. $N_j$ is the total number of $j$-class flares produced in the AR during the 8-day interval, and $\xi_{i}^{\left( j \right)}$ is the magnitude (i.e., digit multipliers from 1.0 to 9.9) of the $j$-class flares. Note that $I_\mathrm{8d}$ measures the AR's total flare productivity during the 8-day interval, and it is different from the daily flare index which is, in our case, $I_\mathrm{8d}$ divided by the time interval 8 days used for counting flares in the AR \citep[e.g.,][]{1996CoSka..26...98A,2005ApJ...629.1141A}.

\section{Results}
\label{s:results}
This section is divided into three parts. In Section~\ref{ss:result_1}, we present the relationship between the mean $\mathrm{\bar{X}}$ of the SHARP parameter time series and the 8-day flare index $I_\mathrm{8d}$ for the 93 major flaring ARs under investigation. In Section~\ref{ss:result_2}, we describe how the detrended fluctuation measurement $\mathrm{\tilde{F}}$ is correlated with $I_\mathrm{8d}$. In Section~\ref{ss:result_3}, we discuss how the relation of $I_\mathrm{8d}$ to $\mathrm{\bar{X}}$ or $\mathrm{\tilde{F}}$ varies if we use an initial portion of the time series for calculation of $\mathrm{\bar{X}}$ and $\mathrm{\tilde{F}}$.

\subsection{Flare Index versus Mean of SHARP Time Series}
\label{ss:result_1}
\begin{figure}
\centering
\includegraphics[width=\textwidth]{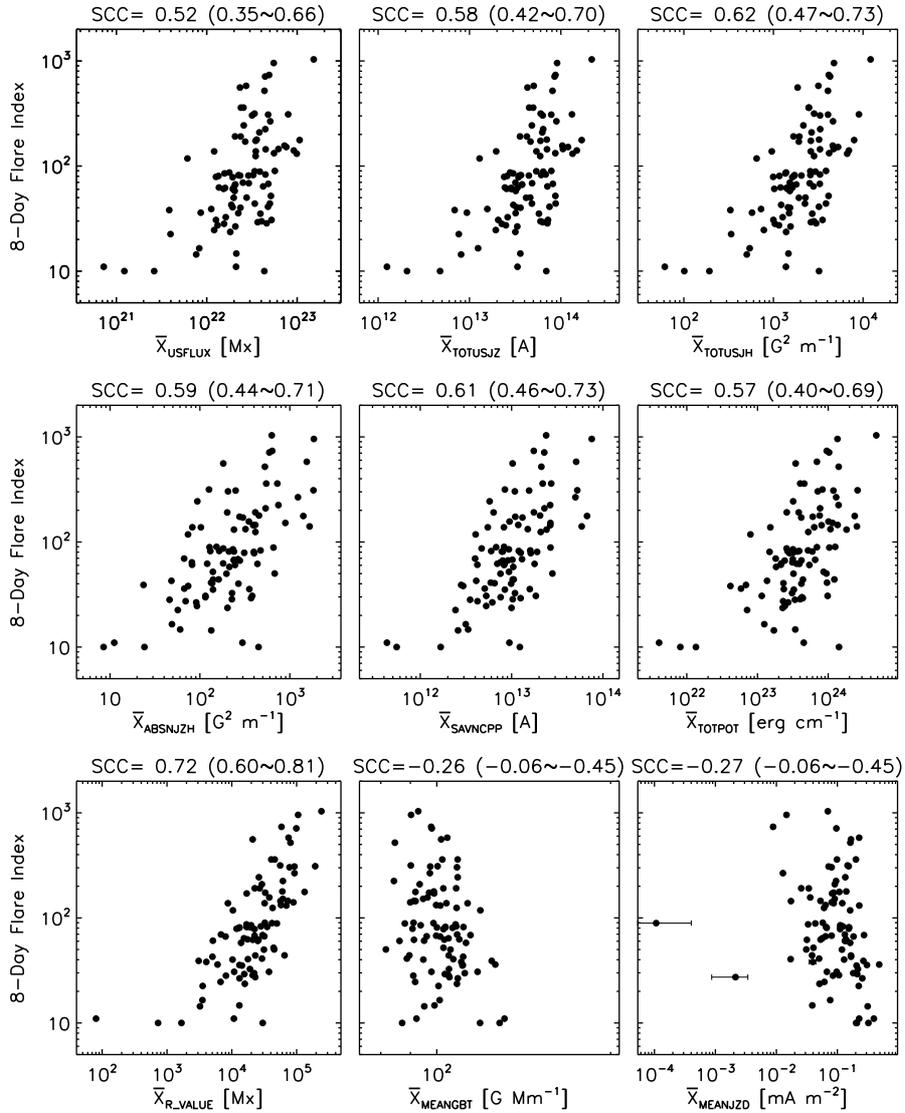}
\caption{Log-log scatter plots of the flare index $I_\mathrm{8d}$ vs. the mean $\mathrm{\bar{X}}$ calculated from 8-day time series of 9 SHARP parameters for 93 major flaring ARs. The Spearman's rank-order correlation coefficient (SCC) and its 95\% confidence interval are denoted in each panel. The uncertainty in calculating $\mathrm{\bar{X}}$ for each AR is also marked with an error bar, but note that in many cases the range of error bars is smaller than the size of data points (shown by circles).}
\label{f2}
\end{figure}

We examine the relationship between $I_\mathrm{8d}$ and $\mathrm{\bar{X}}$ for the 93 major flaring ARs. Figure~\ref{f2} shows $I_\mathrm{8d}$ versus $\mathrm{\bar{X}}$, in log-log space, calculated from the following SHARP parameter time series: (a) the total unsigned magnetic flux USFLUX, (b) the total unsigned vertical current TOTUSJZ, (c) the total unsigned vertical current helicity TOTUSJH, (d) the absolute value of the net vertical current helicity ABSNJZH, (e) the sum of the absolute value of the net current per polarity SAVNCPP, (f) the surface integral of photospheric magnetic free energy density TOTPOT, (g) R\_VALUE, (h) the mean gradient of the total field strength MEANGBT, and (i) the mean vertical current density MEANJZD. The 9 SHARP parameters shown in Figure~\ref{f2} are selected among a total of the 18 parameters for the following reasons: first, USFLUX is considered as reference, and then the others for which $\mathrm{\bar{X}}$ or $\mathrm{\tilde{F}}$ has a better correlation with $I_\mathrm{8d}$, compared to that of USFLUX. The Spearman's rank-order correlation coefficient (SCC) is derived from the log-log scatter plot of $I_\mathrm{8d}$ against $\mathrm{\bar{X}}$, and it is marked in each panel of Figure~\ref{f2} with a 95\% confidence interval using the Fisher's Z-transformation. Note that the SCCs are statistically significant with p-values less than 0.05. In addition, for a given SHARP parameter, we estimate uncertainties in $\mathrm{\bar{X}}$ as follows: (i) $\mathrm{\bar{X}}$ is calculated for each AR, from a set of 100 different time series of the SHARP parameter values in which measurement errors are added; (ii) the uncertainty in $\mathrm{\bar{X}}$ for each AR is then defined as the standard deviation of 100 values of $\mathrm{\bar{X}}$ determined in the previous step (i). The uncertainties are marked in each panel of Figure~\ref{f2}.

First, in the case of USFLUX, we find that for the examined major flaring ARs with the larger $\mathrm{\bar{X}}$ of USFLUX, their 8-day flare productivity tends to be generally higher (SCC=0.52). Note that USFLUX is a well-known parameter that has a moderately good correlation with the occurrence rate and magnitude of solar eruptive events such as flares and CMEs \citep[e.g.,][]{2002ApJ...569.1016F,2002SoPh..209..361T,2003ApJ...595.1296L,2007ApJ...656.1173L}, and it has been frequently used as a reference for the evaluation of any flare-prediction parameter \citep[e.g.,][]{2010ApJ...718...43P,2017SoPh..292..159K}. A better correlation (SCCs of 0.57--0.72) is also found in the parameters that characterize the AR's strong-gradient PILs, current helicity, current and magnetic free energy (i.e., R\_VALUE, TOTUSJH, SAVNCPP, ABSNJZH, TOTUSJZ and TOTPOT, in the order of a higher SCC). Interestingly, these parameters, which have a higher SCC than USFLUX, are calculated from a sum of a physical quantity over an entire AR surface or an extended PIL region (i.e., \textit{extensive} parameters). This is in agreement with previous studies based on AR photospheric magnetic parameters \citep[e.g.,][]{2009ApJ...705..821W,2015ApJ...798..135B}. 

It is also found that $I_\mathrm{8d}$ is weakly anticorrelated with $\mathrm{\bar{X}}$ of the \textit{intensive} SHARP parameters MEANGBT and MEANJZD (SCCs of -0.26 and -0.27, respectively). Note that MEANJZD only has both positive and negative values (i.e., a signed parameter) among the 9 SHARP parameters. Because the mean of the absolute values of MEANJZD during the interval of the entire time series is a measure of the AR's average current imbalance, we further check the relationship of $I_\mathrm{8d}$ with the mean value of the unsigned time series of MEANJZD: as a result, a moderate anticorrelation (SCC=-0.46) is found. The anticorrelation between $I_\mathrm{8d}$ and $\mathrm{\bar{X}}$ of either signed or unsigned MEANJZD may in part be explained from the fact that $\mathrm{\bar{X}}$ of signed or unsigned MEANJZD is anticorrelated with $\mathrm{\bar{X}}$ of USFLUX (i.e., SCC of -0.27 or -0.52, respectively). It seems currents in the examined ARs are fairly neutralized. In that case, the sum of signed values of the vertical current density in the AR's entire photospheric surface may not differ significantly between the ARs of different sizes so that MEANJZD will be in inverse proportion to the size of the ARs. In the case of MEANGBT, the anticorrelation may be inferred as follows. Unlike the vertical magnetic field $\mathrm{B_z}$, the total field strength $\mathrm{B_t}$ does not change its sign across PILs. Thus, MEANGBT will be relatively small in the case of an AR with a highly flare-productive delta sunspot because the strong-field, opposite magnetic polarities are located very close to each other so that the horizontal gradients of $\mathrm{B_t}$ around the AR's PIL will be very small. On the other hand, in the case of a less flare-productive AR with widely separated sunspots, the gradients of $\mathrm{B_t}$ can have relatively large values at the sunspot boundaries. An anticorrelation (SCC=-0.49) between $\mathrm{\bar{X}}$ values of MEANGBT and R\_VALUE can support the inference mentioned above.

In addition, we investigate whether the correlation between $I_\mathrm{8d}$ and $\mathrm{\bar{X}}$ is better than that between $I_\mathrm{8d}$ and any randomly selected, single data point in the time series $\textbf{X}$. For this, a Monte Carlo test is carried out with the time series data of the 9 SHARP parameters shown in Figure~\ref{f2}, as follows: (1) for each of the 93 ARs in the data set, we randomly select a single value from $\textbf{X}$, (2) we calculate the SCC between $I_\mathrm{8d}$ and the randomly selected values, (3) repeating the above two procedures 100 times, we calculate the average of the SCCs. We find that in general $I_\mathrm{8d}$ is slightly better correlated with $\mathrm{\bar{X}}$ than the randomly selected data point at a single point in time (i.e., SCCs are 0.02--0.12 higher). In particular, for the best-correlated parameter, R\_VALUE, the SCC from the log-log plot of $I_\mathrm{8d}$ vs. $\mathrm{\bar{X}}$ is 0.72, while that of $I_\mathrm{8d}$ vs. randomly selected data points is 0.61. Refer to the right two columns in Table~\ref{table_2} for the SCCs calculated from the log-log plots of $I_\mathrm{8d}$ vs. $\mathrm{\bar{X}}$ and $I_\mathrm{8d}$ vs. randomly selected values, respectively, for all of the 9 SHARP parameters.

For the SHARP parameters of which $\mathrm{\bar{X}}$ has a meaningful correlation with $I_\mathrm{8d}$, it is important to understand the degree to which a selected SHARP parameter is related to $I_\mathrm{8d}$, independent of another parameter. We therefore calculate the Pearson partial correlation coefficient between $I_\mathrm{8d}$ and $\mathrm{\bar{X}}$ of one selected among the top 7 most strongly correlated parameters in Figure~\ref{f2} while removing the effect of another well-correlated SHARP parameter. As a result, it is found that the partial correlation coefficients between $I_\mathrm{8d}$ and $\mathrm{\bar{X}}$ of the best-correlated parameter R\_VALUE are in the range of 0.36--0.55 in log-log space, which are still meaningful despite eliminating the effect of one of the other parameters.

\subsection{Flare Index versus Fluctuation of SHARP Time Series}
\label{ss:result_2}
\begin{figure}
\centering
\includegraphics[width=\textwidth]{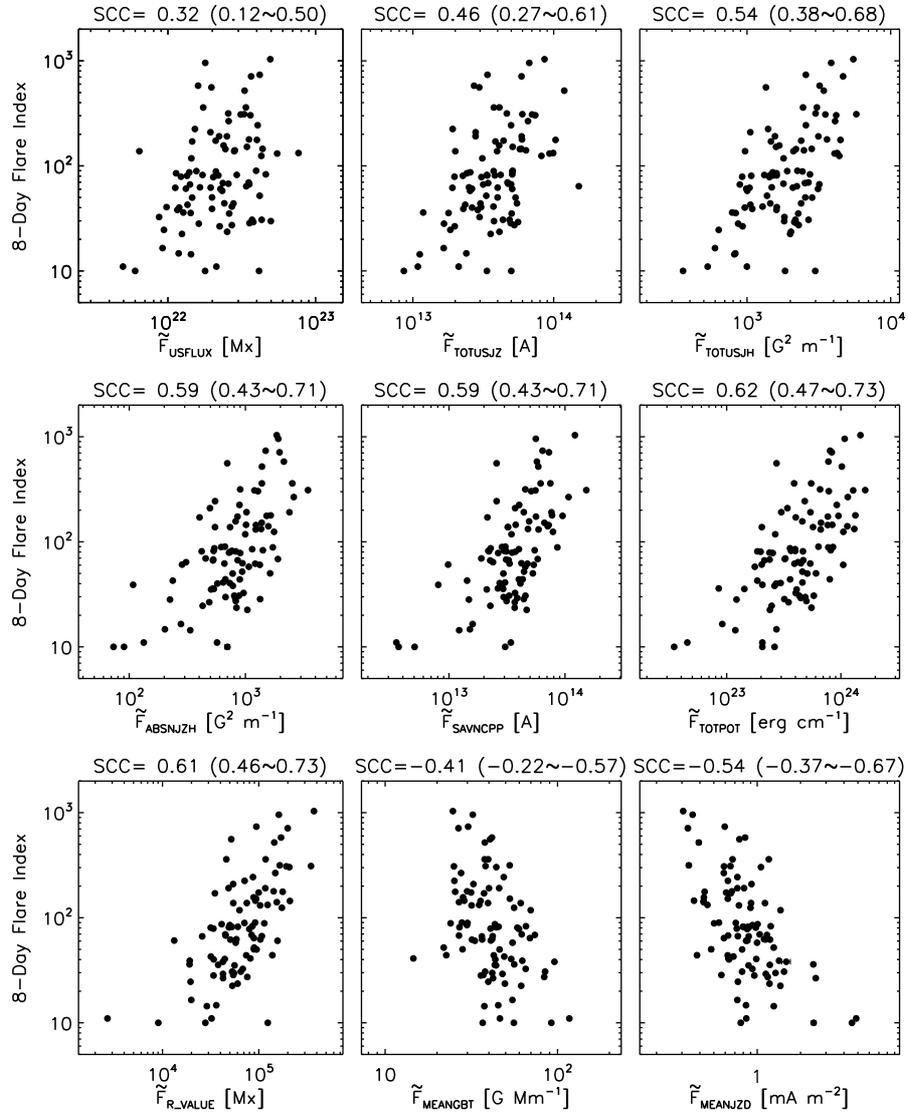}
\caption{Same as in Figure~\ref{f2}, but for $I_\mathrm{8d}$ vs. the detrended fluctuation $\mathrm{\tilde{F}}$.}
\label{f3}
\end{figure}

The relationship between $I_\mathrm{8d}$ and $\mathrm{\tilde{F}}$ is also examined for the 93 major flaring ARs under investigation. Figure~\ref{f3} shows $I_\mathrm{8d}$ as a function of $\mathrm{\tilde{F}}$, in log-log space, determined from the time series of the same SHARP parameters shown in Figure~\ref{f2}. SCCs are marked in Figure~\ref{f3}, and all of them are statistically significant with p-values less than 0.05. The uncertainties in $\mathrm{\tilde{F}}$ for each AR, determined in the same way described in Section~\ref{ss:result_1}, are shown in each panel of Figure~\ref{f3}. For the 9 SHARP parameters, we find that SCCs between $I_\mathrm{8d}$ and $\mathrm{\tilde{F}}$ are in general similar or slightly lower than those between $I_\mathrm{8d}$ and $\mathrm{\bar{X}}$. However, the best SCC of 0.62 is found from the log-log scatter plot of $I_\mathrm{8d}$ vs. $\mathrm{\tilde{F}}$ of the surface integral of photospheric magnetic free energy density TOTPOT, which is much larger than the SCC of 0.32 from that of $I_\mathrm{8d}$ vs. $\mathrm{\tilde{F}}$ of USFLUX. We also find that $I_\mathrm{8d}$ shows a negative correlation with $\mathrm{\tilde{F}}$ of the mean vertical current density MEANJZD (SCC=-0.54) or the mean gradient of the total field strength MEANGBT (SCC=-0.41), respectively. This result suggests that more flare-productive ARs show less fluctuation of MEANJZD and MEANGBT during the time interval of 8 days. To understand the importance of each of the 9 SHARP parameters in relation to $I_\mathrm{8d}$, as in Section~\ref{ss:result_1}, the Pearson partial correlation coefficients are calculated between $I_\mathrm{8d}$ and $\mathrm{\tilde{F}}$. It is found that the partial correlation coefficients between $I_\mathrm{8d}$ and the best-correlated fluctuation parameter TOTPOT are in the range of 0.23--0.59 in log-log space; i.e., the highest partial correlation is obtained with $\mathrm{\tilde{F}}$ of USFLUX as a control variable, while the lowest with $\mathrm{\tilde{F}}$ of R\_VALUE. In addition, in the case that $\mathrm{\tilde{F}}$ of USFLUX is used as a control variable, the partial correlation coefficients remain more or less the same or slightly decrease for most of the parameters, compared to the Pearson correlation coefficients in log-log space without removing the effect of $\mathrm{\tilde{F}}$ of USFLUX.

\begin{figure}
\centering
\includegraphics[width=\textwidth]{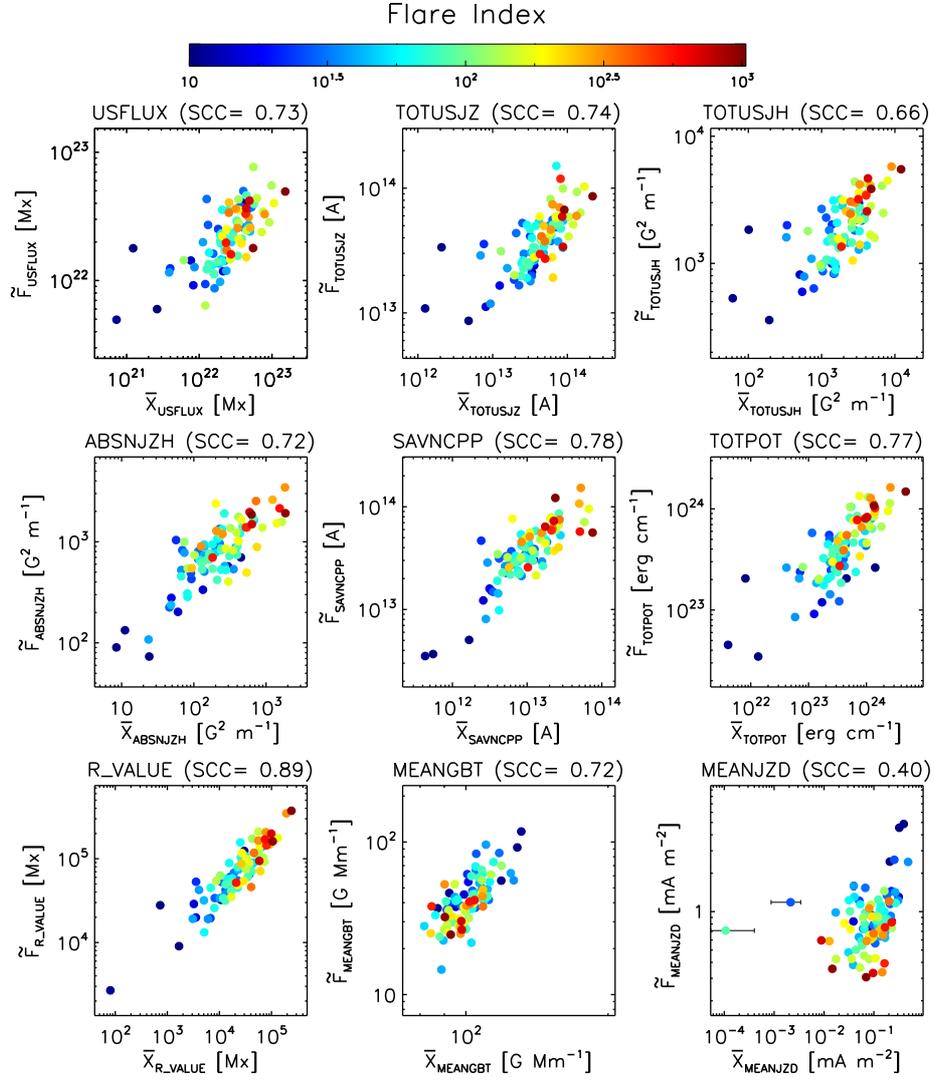}
\caption{Log-log plots of $\mathrm{\tilde{F}}$ vs. $\mathrm{\bar{X}}$ for the 93 major flaring ARs under study. The color of each AR data point indicates $I_\mathrm{8d}$ of the given AR. Refer to the color bar in the top of the figure for the values of $I_\mathrm{8d}$.}
\label{f4}
\end{figure}

It is worthwhile to check whether/how $\mathrm{\tilde{F}}$ is related to $\mathrm{\bar{X}}$ to understand the temporal variation of the SHARP parameters in relation to flare activity. We expect fluctuation in $\textbf{X}$ of a given SHARP parameter to be on the same scale as values in $\textbf{X}$ of the SHARP parameter, suggesting that values of $\mathrm{\tilde{F}}$ will be large or small when values of $\mathrm{\bar{X}}$ are large or small, respectively. We therefore expect $\mathrm{\tilde{F}}$ and $\mathrm{\bar{X}}$ to be significantly correlated, independent of whether or not $\mathrm{\bar{X}}$ is associated flare activity. Figure~\ref{f4} shows log-log plots of $\mathrm{\tilde{F}}$ vs. $\mathrm{\bar{X}}$ calculated from the 93 major flaring ARs for the same SHARP parameters shown in Figures~\ref{f2} and ~\ref{f3}. The color of each AR data point (marked as a circle) indicates the AR's $I_\mathrm{8d}$. From the log-log plots, we find that in general $\mathrm{\bar{X}}$ and $\mathrm{\tilde{F}}$ are well correlated with each other in log-log space (SCCs of 0.66--0.89) even though there is a relatively weak correlation in the case of MEANJZD (SCC=0.40). This suggests that the larger $\mathrm{\bar{X}}$ an AR has for a given magnetic parameter, the larger $\mathrm{\tilde{F}}$ of the parameter it may show during a long-term interval of several days. Investigating the correlations between $\mathrm{\bar{X}}$ of USFLUX and $\mathrm{\tilde{F}}$ of the SHARP parameters, we notice that $\mathrm{\bar{X}}$ of USFLUX shows a positive correlation (SCCs of 0.47--0.73) with $\mathrm{\tilde{F}}$ of some parameters which are positively correlated with $I_\mathrm{8d}$, while it shows a negative correlation with $\mathrm{\tilde{F}}$ of MEANGBT (SCC=-0.63) or MEANJZD (SCC=-0.70) which is anticorrelated with $I_\mathrm{8d}$. It is also found that $I_\mathrm{8d}$ tends to be large in the case of an AR which has small values of both $\mathrm{\bar{X}}$ and $\mathrm{\tilde{F}}$ for MEANGBT and MEANJZD and/or large values for the other parameters.

As explained in Section~\ref{s:data}, $\mathrm{\tilde{F}}$ is a measure of the degree of variation in a given SHARP parameter time series $\textbf{X}$, including not only fluctuations but also linear/non-linear trends. It is defined as the root-mean-square deviation of the cumulative sum with respect to its local trend (i.e., a linear fit). The standard deviation $\sigma$ of $\textbf{X}$ can be also used to quantify the amount of variation or dispersion of $\textbf{X}$. We examine how the relationship between $I_\mathrm{8d}$ and $\mathrm{\tilde{F}}$ is different from that between $I_\mathrm{8d}$ and $\sigma$. SCCs are calculated from the log-log plots of $I_\mathrm{8d}$ vs. $\sigma$ for the 9 SHARP parameters. We find that $I_\mathrm{8d}$ is slightly better correlated with $\mathrm{\tilde{F}}$ than $\sigma$ for most of the parameters. There are a few parameters (e.g., USFLUX and MEANJZD) of which $\sigma$ has a little higher SCC with $I_\mathrm{8d}$ than $\mathrm{\tilde{F}}$; MEANJZD typically does not show a relatively long-term variation pattern in their time series. Refer to the right two columns in Table~\ref{table_3} for the SCCs of $I_\mathrm{8d}$ vs. $\mathrm{\tilde{F}}$ and $I_\mathrm{8d}$ vs. $\sigma$, respectively, for the 9 SHARP parameters.

To account for the expected scaling of $\mathrm{\tilde{F}}$ with $\mathrm{\bar{X}}$, we investigate how normalizing $\mathrm{\tilde{F}}$ by the scale of $\mathrm{\bar{X}}$ affects its relationship with $I_\mathrm{8d}$ as well as with $\mathrm{\bar{X}}$. For this, $\mathrm{\tilde{F}}$ is divided by $\sigma$ of $\textbf{X}$, which indicates the ratio of the detrended fluctuation to $\sigma$. We first check how the rescaled fluctuation parameter $\mathrm{\tilde{F}/\sigma}$ is correlated with $\mathrm{\bar{X}}$ of USFLUX as a reference for the size of ARs. A weak anticorrelation is found between $\mathrm{\tilde{F}/\sigma}$ and $\mathrm{\bar{X}}$ of USFLUX for all the SHARP parameters (SCCs of -0.03 to -0.24) in Figure~\ref{f3} excluding SAVNCPP (SCC=0.03). This may suggest that smaller ARs tend to exhibit greater fractional changes in the parameter values. We also examine the correlation between $I_\mathrm{8d}$ and $\mathrm{\tilde{F}/\sigma}$ for the 9 SHARP parameters; however, a very weak anticorrelation is found between them (SCCs of -0.04 to -0.20) except MEANJZD (SCC=0.11). Because ARs with large values of USFLUX and/or magnetic parameters are not only more flare-productive but they also evolve more dynamically over time, it is reasonable to expect that $I_\mathrm{8d}$ is moderately well correlated with $\mathrm{\bar{X}}$ or $\mathrm{\tilde{F}}$ but only weakly correlated with $\mathrm{\tilde{F}/\sigma}$.

\subsection{Various Time Intervals for Calculation of Mean and Fluctuation}
\label{ss:result_3}
We examine how the correlation between $I_\mathrm{8d}$ and $\mathrm{\bar{X}}$ as well as between $I_\mathrm{8d}$ and $\mathrm{\tilde{F}}$ varies depending on the time interval used for calculation of $\mathrm{\bar{X}}$ and $\mathrm{\tilde{F}}$. SCCs are calculated from the log-log plots of $I_\mathrm{8d}$ vs. $\mathrm{\bar{X}}$ and $I_\mathrm{8d}$ vs. $\mathrm{\tilde{F}}$, of which $\mathrm{\bar{X}}$ and $\mathrm{\tilde{F}}$ are calculated from some different initial portions of the SHARP parameter time series as follows: $\textbf{X}_\mathrm{1d}=\left[\mathrm{X}_{1},\mathrm{X}_{2},\mathrm{X}_{3},\ldots, \mathrm{X}_{120}\right]$, $\textbf{X}_\mathrm{2d}=\left[\mathrm{X}_{1},\mathrm{X}_{2},\mathrm{X}_{3},\ldots, \mathrm{X}_{240}\right]$, $\ldots$, and $\textbf{X}_\mathrm{8d}=\left[\mathrm{X}_{1},\mathrm{X}_{2},\mathrm{X}_{3},\ldots, \mathrm{X}_{960}\right]$. Note that if high SCCs can be found between $I_\mathrm{8d}$ and $\mathrm{\bar{X}}$ or between $I_\mathrm{8d}$ and $\mathrm{\tilde{F}}$ from an initial portion of the time series, then it may help us to estimate a long-term (several days) flare productivity of a given AR only taking into account the measurement of the SHARP parameters for the first few days of the AR's disk passage.

First, as shown in Table~\ref{table_2}, the SCCs between $I_\mathrm{8d}$ and $\mathrm{\bar{X}}$ generally increase as a longer time interval is considered: i.e., using a larger portion of the time series tends to improve the correlation. R\_VALUE shows the best correlation with $I_\mathrm{8d}$ in all the different intervals. It is also interesting that the correlation between $I_\mathrm{8d}$ and $\mathrm{\bar{X}}$, calculated from the time intervals of the first few days in $\textbf{X}$, is slightly worse than that between $I_\mathrm{8d}$ and a randomly selected single data point in $\textbf{X}$, for most of the 9 SHARP parameters except the poorly anticorrelated parameters MEANJZD and MEANGBT. We speculate this is due to the fact that $\mathrm{\bar{X}}$ calculated from the early, short-term phase of $\textbf{X}$ may not represent the magnetic field properties of the major flaring ARs which in general dynamically evolve during their 8-day disk passage.

With respect to $\mathrm{\tilde{F}}$, a similar increasing trend of the SCCs between $I_\mathrm{8d}$ and $\mathrm{\tilde{F}}$, as found between $I_\mathrm{8d}$ and $\mathrm{\bar{X}}$, is also found as the time interval considered for calculation of $\mathrm{\tilde{F}}$ gets longer. Refer to Table~\ref{table_3} for the details of the SCCs for the 9 SHARP parameters. $I_\mathrm{8d}$ shows the highest correlation (SCCs of 0.51--0.65 with $\mathrm{\tilde{F}}$ of the photospheric magnetic free energy density parameter TOTPOT for all the intervals except for the 2-day interval). We also find that a relatively high SCC of 0.61 can be obtained between $I_\mathrm{8d}$ and $\mathrm{\tilde{F}}$ of the best-correlated parameter TOTPOT calculated from the 4-day interval (i.e., using the half of the time series), compared to that (SCC=0.62) calculated from the 8-day interval.

\section{Summary and Discussions}
\label{s:sum}
In this study, we have investigated the relationship of the long-term (i.e., 8-day) flare index $I_\mathrm{8d}$ with the mean $\mathrm{\bar{X}}$ or detrended fluctuation $\mathrm{\tilde{F}}$ calculated from 8-day time series of 18 photospheric magnetic parameters (called SHARP parameters) for 93 major flaring ARs. Our findings can be summarized as follows:
\begin{enumerate}[(i)]
\item{A meaningful correlation is found between $I_\mathrm{8d}$ and $\mathrm{\bar{X}}$ as well as between $I_\mathrm{8d}$ and $\mathrm{\tilde{F}}$ with SCCs of $\sim$0.6--0.7 for some SHARP parameters. In the case of $\mathrm{\bar{X}}$, $I_\mathrm{8d}$ shows its best correlation with the PIL-related parameter R\_VALUE (SCC=0.72), while with $\mathrm{\tilde{F}}$ of the surface integral of magnetic free energy density TOTPOT (SCC=0.62). For many of the SHARP parameters, the SCCs between $I_\mathrm{8d}$ and $\mathrm{\bar{X}}$ are slightly higher than those between $I_\mathrm{8d}$ and $\mathrm{\tilde{F}}$.}
\item{Using only an initial portion (e.g., first 4 days) of 8-day time series of some SHARP parameters, a similar correlation of $I_\mathrm{8d}$ with $\mathrm{\bar{X}}$ or with $\mathrm{\tilde{F}}$ can be obtained as found in (i) using the entire time series.}
\end{enumerate}

Several studies have been carried out to find out photospheric magnetic parameters which can be used to understand flare productivity of ARs, in particular, with respect to flare occurrence in the next short-term (e.g., 24-hr) interval following the parameter measurement. For example, the total unsigned magnetic flux parameter USFLUX, which is one of the most frequently examined parameters and measures the size of ARs, shows that it has a moderately good correlation with soft X-ray flare index of ARs \citep[e.g.,][]{2010ApJ...720..717A}, as well as a somewhat flare-predictive capability \citep[e.g.,][]{2007ApJ...656.1173L,2010ApJ...718...43P,2017SoPh..292..159K}. In addition, compared to USFLUX, similarly good or (slightly) better flare-predictive parameters were reported, such as the mean photospheric magnetic free energy density proxy MEANPOT \citep[e.g.,][]{2012SoPh..280..165Y,2013ApJ...774L..27Y}, the total unsigned vertical current TOTUSJZ \citep[e.g.,][]{2007ApJ...656.1173L} and the total unsigned vertical current helicity TOTUSJH \citep[e.g.,][]{2015ApJ...798..135B}. As already reported in those previous studies, we also found that $\mathrm{\bar{X}}$ of USFLUX shows a moderately good correlation with the long-term flare productivity (i.e., $I_\mathrm{8d}$). Moreover, $\mathrm{\bar{X}}$ of some parameters that quantitatively measure the AR's morphological complexity, non-potentiality or magnetic free energy shows a relatively better correlation than USFLUX. An even more remarkable finding is that $\mathrm{\tilde{F}}$ of those non-potentiality-associated parameters shows a comparably good correlation with the flare productivity. Here we report this for the first time examining the SHARP parameter time series of the 93 major flaring ARs. It is however not clear how $\mathrm{\tilde{F}}$ is related to the AR's flaring activity. We speculate that short-term (tens minutes to a few hours), large variations of some SHARP magnetic parameters under study, indicated by $\mathrm{\tilde{F}}$, may (partly) represent how unstable the AR's magnetic system is and/or whether a flare will be triggered in relation to a rapid change or increase of twists, currents, and so on. $\mathrm{\tilde{F}}$ also captures random variations in $\textbf{X}$, and these fluctuations may be relevant for the evolution of an AR into a critical state in which a small perturbation can trigger a flare, as suggested by SOC models \citep[e.g., refer to][and references therein]{2012A&A...539A...2A}. However, the fact that $\mathrm{\tilde{F}}$ is not much more significant for flaring than $\mathrm{\bar{X}}$ may suggest that the corona may not be predominantly driven to flare by random magnetic evolution at the photosphere, as would be expected based upon SOC models. On the other hand, $\mathrm{\tilde{F}}$ may measure some changes in the AR's magnetic field configuration and associated currents in the course of a flare and its corresponding eruption, if any \citep[e.g.,][]{2012ApJ...745L..17W, 2012ApJ...757L...5W, 2013SoPh..287..415P, 2014ApJ...788...60J, 2014ApJ...782L..31W}.

The results from this study suggest that for a given AR, examining its $\mathrm{\bar{X}}$ and $\mathrm{\tilde{F}}$ from time series with a few days of multiple photospheric (vector) magnetic parameters, we may be able to estimate the AR's long-term flare productivity. Further studies on SHARP parameter time series data, using the same analysis of $\mathrm{\bar{X}}$ and $\mathrm{\tilde{F}}$ but with an extended data set including less flare-productive ARs (e.g., ARs that produce C-class flares only), will help us to understand whether the same relation of the long-term flare index with $\mathrm{\bar{X}}$ or $\mathrm{\tilde{F}}$ still exists or not. In the case of flare-quiet ARs, it will be also interesting to check whether there is any threshold-like value of $\mathrm{\bar{X}}$ or $\mathrm{\tilde{F}}$ for some SHARP parameters between flaring and flare-quiet ARs. Furthermore, a different parameterization of SHARP parameter time series, such as linear trends (increase or decrease), can be applied to find any characteristic variation in magnetic parameters before major flares. We expect these kinds of parameter-based studies will improve our understanding of which AR magnetic parameter(s) can provide some useful information regarding the flare energy build-up and triggering processes.

\begin{acks}
The authors would like to thank the anonymous referee for many constructive comments. This work was supported by the BK21 plus program through the National Research Foundation (NRF) funded by the Ministry of Education of Korea, the Basic Science Research Program through the NRF funded by the Ministry of Education (NRF-2016R1A2B4013131), NRF of Korea Grant funded by the Korean Government (NRF-2013M1A3A3A02042232), the Korea Astronomy and Space Science Institute under the R\&D program supervised by the Ministry of Science, ICT and Future Planning, the Korea Astronomy and Space Science Institute under the R\&D program ‘Development of a Solar Coronagraph on International Space Station (Project No. 2017-1-851-00)’ supervised by the Ministry of Science, ICT and Future Planning, and Institute for Information \& communications Technology Promotion (IITP) grant funded by the Korea government (MSIP) (2018-0-01422, Study on analysis and prediction technique of solar flares). The data used in this work are courtesy of the NASA/SDO and HMI science teams, as well as the Geostationary Satellite System (GOES) team. This research has made use of NASA’s Astrophysics Data System (ADS). S.-H.P. was supported by the European Union Horizon 2020 research and innovation programme under grant agreement No.~640216 (FLARECAST; \url{http://flarecast.eu}) and by MEXT/JSPS KAKENHI Grant No.~JP15H05814.
\end{acks}
\hfill \break
\hfill \break
\noindent \textbf{Disclosure of Potential Conflicts of Interest} The authors declare that they have no conflicts of interest.

\bibliographystyle{spr-mp-sola.bst} 
\bibliography{ref.bib}

\clearpage

\begin{table}
\begin{adjustbox}{width=\textwidth, center=\textwidth}
\caption{Space-weather HMI Active Region Patch (SHARP) parameters$^\ast$}
\label{table_1}
\begin{tabular}{clll}
\hline
Keyword   & Description                                        & Units    & formula  \\ 
\hline
USFLUX    & Total unsigned magnetic flux                                & Mx       & $\Phi = \sum \left| B_z \right| dA$  \\
MEANGAM   & Mean angle of field from radial                    & Degrees  & $\overline{\gamma} = \frac{1}{N} \sum \arctan \left( \frac{B_h}{B_z} \right)$  \\
MEANGBT   & Mean gradient of total field                       & G Mm$^{-1}$   & $\overline{\left| \nabla B_{tot} \right|} = \frac{1}{N} \sum \sqrt{\left( \frac{\partial B}{\partial x} \right)^2 + \left( \frac{\partial B}{\partial y} \right)^2}$  \\
MEANGBZ   & Mean gradient of vertical field                    & G Mm$^{-1}$   & $\overline{\left| \nabla B_z \right|} = \frac{1}{N} \sum \sqrt{\left( \frac{\partial B_z}{\partial x} \right)^2 + \left( \frac{\partial B_z}{\partial y} \right)^2}$  \\
MEANGBH   & Mean gradient of horizontal field                  & G Mm$^{-1}$   & $\overline{\left| \nabla B_h \right|} = \frac{1}{N} \sum \sqrt{\left( \frac{\partial B_h}{\partial x} \right)^2 + \left( \frac{\partial B_h}{\partial y} \right)^2}$  \\
MEANJZD   & Mean vertical current density                      & mA m$^{-2}$   & $\overline{J_z} \propto \frac{1}{N} \sum \left( \frac{\partial B_y}{\partial x} - \frac{\partial B_x}{\partial y} \right)$  \\
TOTUSJZ   & Total unsigned vertical current                    & A        & $J_{z_{total}} = \sum \left| J_z \right| dA$  \\
MEANALP   & Mean characteristic twist parameter, $\alpha$      & Mm$^{-1}$     & $\alpha_{total} \propto \frac{\sum J_z \cdot B_z}{\sum B_z^2}$  \\
MEANJZH   & Mean vertical current helicity                     & G$^2$ m$^{-1}$   & $\overline{H_c} \propto \frac{1}{N} \sum B_z \cdot J_z$  \\
TOTUSJH   & Total unsigned vertical current helicity           & G$^2$ m$^{-1}$   & $H_{c_{total}} \propto \sum \left| B_z \cdot J_z \right|$  \\
ABSNJZH   & Absolute value of the net vertical current helicity & G$^2$ m$^{-1}$   & $H_{c_{abs}} \propto \left| \sum B_z \cdot J_z \right|$  \\
SAVNCPP   & Sum of the absolute value of the net current per polarity & A        & $J_{z_{sum}} \propto \left| \sum^{B_z^+} J_z dA \right| + \left| \sum^{B_z^-} J_z dA \right|$  \\
MEANPOT   & Mean photoshperic magnetic free energy density     & erg cm$^{-3}$ & $\overline{\rho} \propto \frac{1}{N} \sum \left( \mathbf{B}^{Obs} - \mathbf{B}^{Pot} \right)^2$  \\
TOTPOT    & Surface integral of photospheric magnetic free energy density    & erg cm$^{-1}$ & $\rho_{tot} \propto \sum \left( \mathbf{B}^{Obs} - \mathbf{B}^{Pot} \right)^2 dA$  \\
MEANSHR   & Mean shear angle                                   & Degrees  & $\overline{\Gamma} = \frac{1}{N} \sum \arccos \left( \frac{\mathbf{B}^{Obs} \cdot \mathbf{B}^{Pot}}{\left| B^{Obs} \right| \left| B^{Pot} \right|} \right)$  \\
SHRGT45   & Fraction of Area with shear > 45$^\circ$       &          & Area with shear > 45$^\circ$ / total area  \\
R\_VALUE  & Sum of flux near polarity inversion line           & Mx       & $\Phi = \sum \left| B_{LoS} \right| dA$ within $R$ mask  \\
AREA\_ACR & Area of strong field pixels in the active region   &          & Area = $\sum \text{Pixels}$ \\
\hline
\multicolumn{4}{l}{$^\ast$ Further description of the SHARP parameters can be found in \citet{2015ApJ...798..135B} and references therein.}
\end{tabular}
\end{adjustbox}
\end{table}

\begin{table}
\begin{adjustbox}{width=\textwidth, center=\textwidth}
\centering
\caption{The Spearman's rank-order correlation coefficients (SCCs) calculated from log-log\\\hspace{\textwidth} plots of 8-day flare index $I_\mathrm{8d}$ vs. the mean values of time series of 9 SHARP parameters with\\\hspace{\textwidth} different intervals, starting from the first day to the entire 8 days. The highest SCC\\\hspace{\textwidth} at each interval is marked in bold italics. SCCs of $I_\mathrm{8d}$ with a randomly selected, single data\\\hspace{\textwidth} point in the 8-day time series are also shown for comparison.}
\label{table_2}
\begin{tabular}{crrrrrrrrc}
\hline
SHARP      & \multirow{2}{*}{1-Day} & \multirow{2}{*}{2-Day} & \multirow{2}{*}{3-Day} & \multirow{2}{*}{4-Day} & \multirow{2}{*}{5-Day} & \multirow{2}{*}{6-Day} & \multirow{2}{*}{7-Day} & \multirow{2}{*}{8-Day} & Random \\
Parameters &  &  &  &  &  &  &  &  & Point\\
\hline
USFLUX   & 0.40  & 0.42  & 0.45  & 0.48  & 0.49  & 0.50  & 0.52  & 0.52  & 0.50  \\
TOTUSJZ  & 0.45  & 0.48  & 0.50  & 0.53  & 0.55  & 0.56  & 0.57  & 0.58  & 0.54  \\
TOTUSJH  & 0.48  & 0.50  & 0.53  & 0.57  & 0.59  & 0.60  & 0.61  & 0.62  & 0.57  \\
ABSNJZH  & 0.44  & 0.47  & 0.50  & 0.52  & 0.56  & 0.58  & 0.60  & 0.59  & 0.48  \\
SAVNCPP  & 0.40  & 0.46  & 0.46  & 0.51  & 0.56  & 0.60  & 0.62  & 0.61  & 0.49  \\
TOTPOT   & 0.42  & 0.44  & 0.48  & 0.50  & 0.52  & 0.54  & 0.56  & 0.57  & 0.52  \\
R\_VALUE & {\textbf{\textit{0.54}}}  & {\textbf{\textit{0.57}}}  & {\textbf{\textit{0.61}}}  & {\textbf{\textit{0.65}}}  & {\textbf{\textit{0.68}}}  & {\textbf{\textit{0.71}}}  & {\textbf{\textit{0.72}}}  & {\textbf{\textit{0.72}}}  & {\textbf{\textit{0.61}}}  \\
MEANGBT  & -0.19 & -0.21 & -0.22 & -0.23 & -0.22 & -0.23 & -0.25 & -0.26 & -0.18 \\
MEANJZD  & -0.23 & -0.22 & -0.26 & -0.21 & -0.29 & -0.29 & -0.29 & -0.27 & -0.20 \\
\hline
\end{tabular}
\end{adjustbox}
\end{table}

\begin{table}
\begin{adjustbox}{width=\textwidth, center=\textwidth}
\caption{Same as Table 2, but for SCCs between $I_\mathrm{8d}$ and the detrended fluctuation values of\\\hspace{\textwidth} time series of 9 SHARP parameters. SCCs of $I_\mathrm{8d}$ with the standard deviation of the times series\\\hspace{\textwidth} are also shown for reference.}
\label{table_3}
\begin{tabular}{crrrrrrrrc}
\hline
SHARP      & \multirow{2}{*}{1-Day} & \multirow{2}{*}{2-Day} & \multirow{2}{*}{3-Day} & \multirow{2}{*}{4-Day} & \multirow{2}{*}{5-Day} & \multirow{2}{*}{6-Day} & \multirow{2}{*}{7-Day} & \multirow{2}{*}{8-Day} & Standard \\
Parameters &  &  &  &  &  &  &  &  & Deviation \\
\hline
USFLUX   & 0.23  & 0.19  & 0.24  & 0.31  & 0.32  & 0.35  & 0.35  & 0.32  & 0.32  \\
TOTUSJZ  & 0.47  & 0.43  & 0.42  & 0.50  & 0.46  & 0.48  & 0.48  & 0.46  & 0.45  \\
TOTUSJH  & 0.47  & 0.47  & 0.51  & 0.57  & 0.58  & 0.59  & 0.59  & 0.54  & 0.51  \\
ABSNJZH  & 0.53  & {\textbf{\textit{0.55}}}  & 0.55  & 0.57  & 0.57  & 0.59  & 0.62  & 0.59  & 0.58  \\
SAVNCPP  & 0.37  & 0.47  & 0.47  & 0.51  & 0.53  & 0.55  & 0.61  & 0.59  & 0.58  \\
TOTPOT   & {\textbf{\textit{0.53}}}  & 0.51  & {\textbf{\textit{0.57}}}  & {\textbf{\textit{0.61}}}  & {\textbf{\textit{0.60}}}  & {\textbf{\textit{0.63}}}  & {\textbf{\textit{0.65}}}  & {\textbf{\textit{0.62}}}  & 0.59  \\
R\_VALUE & 0.50  & 0.50  & 0.55  & 0.59  & 0.59  & 0.60  & 0.61  & 0.61  & {\textbf{\textit{0.61}}}  \\
MEANGBT  & -0.18 & -0.27 & -0.31 & -0.37 & -0.38 & -0.43 & -0.44 & -0.41 & -0.36 \\
MEANJZD  & -0.24 & -0.29 & -0.39 & -0.46 & -0.47 & -0.48 & -0.52 & -0.54 & -0.61 \\
\hline
\end{tabular}
\end{adjustbox}
\end{table}

\end{article} 
\end{document}